\documentclass[aps,prl,twocolumn,showpacs,preprintnumbers,amsmath,amssymb,superscriptaddress]{revtex4-2}
\usepackage{epsfig,amsopn}
\usepackage{graphicx}
\usepackage{epstopdf}
\usepackage{sidecap}
\usepackage{color}
\usepackage{amsmath,amssymb}
\usepackage{amsfonts}
\usepackage{textcomp}
\usepackage{enumerate}
\usepackage{soul}
\usepackage{float}
\usepackage{scalerel}
\usepackage{braket}
\usepackage{tikz}
\usepackage{natbib}
\usepackage{booktabs}
\usepackage{comment}
\usepackage[ colorlinks = true,
linkcolor = red,
urlcolor  = red,
citecolor = blue,
anchorcolor = red]{hyperref}

\newcommand{\beq}{\begin{equation}}
  \newcommand{\eeq}{\end{equation}}
\newcommand{\bqa}{\begin{eqnarray}}
  \newcommand{\eqa}{\end{eqnarray}}

\newcommand{\moire}{moir$\mathrm{\acute{e}}\;$}

\begin{document}
	\title{Hidden $f$-wave TIVC and Quantum Phase Transitions in Metallic States of Twisted Bilayer Graphene}
	  \author{Arindam Pramanik}
        \email{apramanik@agh.edu.pl}
        \affiliation{Department of Theoretical Physics, Tata Institute of Fundamental Research,
          Homi Bhabha Road, Mumbai 400005, India }
          \affiliation{Academic Centre for Materials and Nanotechnology, AGH University of Krakow, Al. Mickiewicza 30, 30-059 Krakow, Poland}
          \author{Unmesh Ghorai}
          \affiliation{Department of Theoretical Physics, Tata Institute of Fundamental Research,
          Homi Bhabha Road, Mumbai 400005, India }
          \affiliation{School of Physics and Astronomy, Tel Aviv University, Tel Aviv 69978, Israel}
	\author{Rajdeep Sensarma}
	\affiliation{Department of Theoretical Physics, Tata Institute of Fundamental Research,
		Homi Bhabha Road, Mumbai 400005, India }
	%
	%
	%
	\begin{abstract}
		We study the ground state phase diagram of twisted bilayer graphene at magic angle as function of carrier density and the Coulomb interaction. The carrier density is varied between the fillings $\nu=-2$ and $\nu=-3$ and the Coulomb interaction is tuned by changing the gate distance. Our results reveal that the well established  Kramers intervalley coherent (KIVC) phase near $\nu=-2$, contains a hidden $f-$wave time-reversal symmetric intervalley coherent (TIVC) order, whose strength gradually increases within the metallic regime of the intervalley coherent (IVC) phase away from $\nu=-2$. This finding is interesting in the light of a recent experimental work \cite{Nuckolls_Nature_2023}, where the signatures of the TIVC phase has been reported in ultra-clean samples. We further show that the system goes through multiple phase transitions as the carrier density is tuned. Near the $\nu=-2$, the IVC phase is the ground state, which gives way to valley polarized phase through a second-order-like phase transition as the carrier density is lowered. Further tuning of density pushes the system into the quantum anomalous Hall phase through gradual development of the sublattice polarization. These phase transitions are accompanied by major reconstructions of the Fermi contours.
	\end{abstract}

	\maketitle
        
 \section{Introduction}
Magic-angle twisted bilayer graphene (MATBG) has emerged as a remarkable platform for studying strongly correlated physics in two-dimensional materials. The formation of nearly flat electronic bands at small twist angles dramatically suppresses the kinetic energy \cite{Trambly_NanoLett_2010, Morell_PRB_2010, Bistritzer_PNAS_2011}, allowing electron-electron (e-e) interactions to play a dominant role in determining the electronic properties of the system. These flat bands, arising from the spin, valley, and band degrees of freedom, give rise to an eight-band manifold that hosts a rich variety of interaction-driven phases. Experiments have revealed correlated insulating states \cite{Cao_CI_Nature_2018, Nuckolls_Nature_2020, Cao_SC_Nature_2018, Yankowitz_Science_2019, Lu_Nature_2019}, superconductivity \cite{Cao_SC_Nature_2018, Yankowitz_Science_2019, Lu_Nature_2019, Oh_Nature_2021, Saito_NatPhys_2020, Arora_Nature_2020, Stepanov_Nature_2020, Liu_Science_2021}, ferromagnetism \cite{Sharpe_Science_2019, Sherlin_Science_2020}, strange-metal behaviour \cite{Cao_SM_PRL_2020}, and nematic order \cite{Kerelsky_Nature_2019, Cao_Nematic_Science_2021, Choi_NatPhys_2019} as the carrier density, displacement field, and other external parameters are varied. The coexistence and competition among these phases, together with their sensitivity to sample conditions, make MATBG a particularly rich platform for investigating the interplay between electronic correlations, symmetry breaking, and unconventional superconductivity, as highlighted by extensive experimental and theoretical investigations \cite{Andrei_NatMater_2020, Balents_NatPhys_2020, Ledwith_AnnPhys_2021, Bhowmik_RepProgPhys_2024, Nuckolls_NatRevMater_2024, Yang_NatlSciRev_2026}.

A particularly interesting regime occurs between the correlated insulating states at filling factors $\nu=-2$ and $\nu=-3$, where $\nu$ denotes the number of electrons relative to charge neutrality per moir\'e unit cell. The eight flat bands accommodate eight electrons per moir\'e unit cell, with $\nu=-4$ and $\nu=+4$ corresponding to the empty and completely filled flat-band manifolds, respectively. At $\nu=-2$, numerous experiments have reported a robust correlated insulating state \cite{Cao_CI_Nature_2018, Yankowitz_Science_2019, Lu_Nature_2019}, whose nature has recently been identified as an intervalley coherent (IVC) state through scanning tunnelling spectroscopy \cite{Nuckolls_Nature_2023}. At $\nu=-3$, another correlated insulating state emerges, but its nature depends on the experimental setting. When MATBG is aligned with an hBN substrate \cite{Sherlin_Science_2020} or when an external magnetic field is applied \cite{Choi_Nature_2021, Saito_NatPhys_2021, Park_Nature_2021, Das_PRL_2022}, the insulator is a quantum anomalous Hall (QAH) state, whereas in unaligned, zero-field samples no QAH behaviour has been reported \cite{Yankowitz_Science_2019, Lu_Nature_2019}. Although the QAH state was observed at $\nu=+3$, the many-body particle--hole symmetry of the flat-band-projected Hamiltonian maps $\nu\rightarrow-\nu$
\cite{Bernevig_PRB_2021, Xie_PRB_2023}, so the $\nu=-3$ state is the particle--hole conjugate of the observed one. Between these two fillings, superconductivity has been observed in several experiments \cite{Cao_SC_Nature_2018, Yankowitz_Science_2019, Lu_Nature_2019, Oh_Nature_2021, Saito_NatPhys_2020, Arora_Nature_2020, Stepanov_Nature_2020, Liu_Science_2021}, although the precise location and extent of the superconducting region vary between samples. Thus, the region connecting the IVC insulator at $\nu=-2$ and the insulating state at $\nu=-3$ provides an opportunity to investigate the competition and evolution of distinct symmetry-broken phases in MATBG.

Despite extensive theoretical efforts \cite{Po_PRX_2018, Xu_PRL_2018, Isobe_PRX_2018, Lian_PRL_2019, You_QuantumMater_2019, Guo_PRB_2018, Kennes_PRB_2018, Wu_PRL_2018, Peltonen_PRB_2018, Chichinadze_PRB_2020, Shavit_PRL_2021}, the microscopic origin of superconductivity in MATBG remains an open question. Although superconductivity frequently emerges in the vicinity of correlated insulating phases, its occurrence does not appear to require the presence of a well-developed insulating state \cite{Saito_NatPhys_2020,Stepanov_Nature_2020}. Scanning tunnelling measurements have provided evidence for unconventional superconducting behaviour \cite{Oh_Nature_2021}, while other experiments suggest that superconductivity can persist independently of the correlated insulating phases \cite{Saito_NatPhys_2020, Arora_Nature_2020}. In particular, superconductivity survives over a broader detuning of twist angles than the insulating states and can even be enhanced when the latter are suppressed by increasing the dielectric screening of e-e interactions \cite{Saito_NatPhys_2020,Arora_Nature_2020, Stepanov_Nature_2020, Liu_Science_2021}. Such observations point towards a competition between superconductivity and correlated insulating orders, suggesting that superconductivity may arise from an instability of the underlying normal state rather than simply from doping an insulating phase. This perspective highlights the importance of understanding the nature of the normal state, particularly its symmetry-broken phases and their evolution with carrier density, to gain insight into the mechanisms that may give rise to superconductivity in MATBG. A comprehensive understanding of how the competing symmetry-broken orders evolve with carrier density in the metallic regime ($-3 < \nu <-2$) and interaction strength, and how this evolution is reflected in the electronic structure, remains an important open problem.

In this work, we address these questions using Hartree-Fock (HF) theory to investigate the interaction-dominated normal-state phases of MATBG between $\nu=-2$ and $\nu=-3$. The correlated phase space of MATBG have been extensively investigated previously using a variety of theoretical approaches, including strong-coupling theories \cite{Kang_PRL_2019, Seo_PRL_2019, Bultinck_PRX_2020, Lian_PRB_2021}, HF theory \cite{Fuchun_PRB_2020, Xie_PRL_2020, Hejazi_PRR_2021, Liu_PRR_2021, Liu_PRB_2021, Kwan_PRX_2021, Zhang_PRL_2022, Xie_PRB_2023, Kwan_PRB_2024}, and several other numerical methods such as density matrix renormalization group \cite{Kang_PRB_2020, Wang_PRB_2023, Soejima_PRB_2020}, exact diagonalization \cite{Potasz_PRL_2021, Xie_PRB_2021}, Monte Carlo \cite{Hofmann_PRX_2022, Liao_PRX_2021, Zhang_CPL_2021, Xiao_PRR_2025, Huang_NatCommun_2025}, dynamical mean-field theory \cite{Datta_NatCommun_2023, Rai_PRX_2024}, and coupled cluster approach \cite{Faulstich_PRB_2023}.  We examine how the competition between different symmetry-broken orders in the focused metallic regime is influenced by the strength of the e-e interaction, which we tune through the distance ($d$) to the metallic gates. Our analysis reveals a rich sequence of phases connecting the IVC and QAH insulating states, with the nature of the intermediate phases and the transitions between them depending sensitively on the interaction strength. At $\nu=-2$, the dominant ground-state order for unstrained MATBG is a spin-polarized Kramers intervalley coherent (KIVC) phase, consistent with previous theoretical studies \cite{Bultinck_PRX_2020, Wagner_PRL_2022,
Lian_PRB_2021}. Beyond this dominant order, however, the individual quasiparticle bands exhibit a richer internal structure involving $f$-wave valley polarization (VP) and time-reversal-symmetric intervalley coherence (TIVC) order, in addition to the conventional $s$-wave KIVC order. Although these additional orders largely cancel between the occupied quasiparticle bands at $\nu=-2$, their contributions become increasingly important upon doping. In particular, the $f$-wave TIVC component grows continuously in the metallic regime and reaches a substantial fraction of the dominant KIVC order near the boundary of its stability. This provides a possible route to stabilizing TIVC order even in the absence of electron-phonon coupling. This finding is particularly interesting in light of experimental observations of TIVC signatures in ultra-low-strain samples \cite{Nuckolls_Nature_2023} and theoretical proposals that electron-phonon interactions may favour this order \cite{Kwan_PRB_2024}.

Upon further hole doping, the system undergoes a sequence of transitions involving valley polarization and QAH order. The IVC phase gives way to a VP state near $\nu\sim-2.48$ through a first-order-like transition. The subsequent evolution depends strongly on the interaction strength. For relatively strong interactions, corresponding to larger gate distances, sublattice polarization develops on top of the existing valley polarization, driving the system towards a QAH phase through a continuous transition. For weaker interactions, the VP state remains stable over a wider filling range before giving way to a mixed phase involving KIVC, TIVC, and QAH orders. A robust QAH phase extending to $\nu=-3$ is obtained only for sufficiently strong interactions. This result highlights an important aspect of the phase diagram: even in the absence of hBN alignment, which explicitly breaks sublattice symmetry and favours QAH order, sufficiently strong e-e interactions can stabilize a QAH ground state. The competition between valley and sublattice polarization therefore provides a natural framework for understanding the sensitivity of the $\nu=-3$ insulating state to the interaction strength and experimental environment.

Finally, we investigate how the evolution of the symmetry-broken phases is reflected in the quasiparticle band structure. In particular, we examine the fate of the Dirac nodes at $K_M$ and $K'_M$ of the moir\'e Brillouin zone (mBZ) and establish how their gapping is connected to the underlying symmetry breaking. Depending on the interaction strength, the gap opening is associated predominantly with either the spontaneous breaking of $C_3$ symmetry or the development of sublattice polarization. We further investigate the reconstruction of the quasiparticle bandwidths and Fermi contours across the phase transitions, identifying the associated Lifshitz transitions. These results establish a connection between the evolution of competing orders, the symmetries of the interacting ground state, and the reconstruction of the electronic structure, providing a unified picture of the normal-state phases of MATBG between $\nu=-2$ and $\nu=-3$.

\section{Model and Method}

\subsection{Noninteracting model}

\begin{figure*}[t]
\begin{center}
\includegraphics[scale=.8]{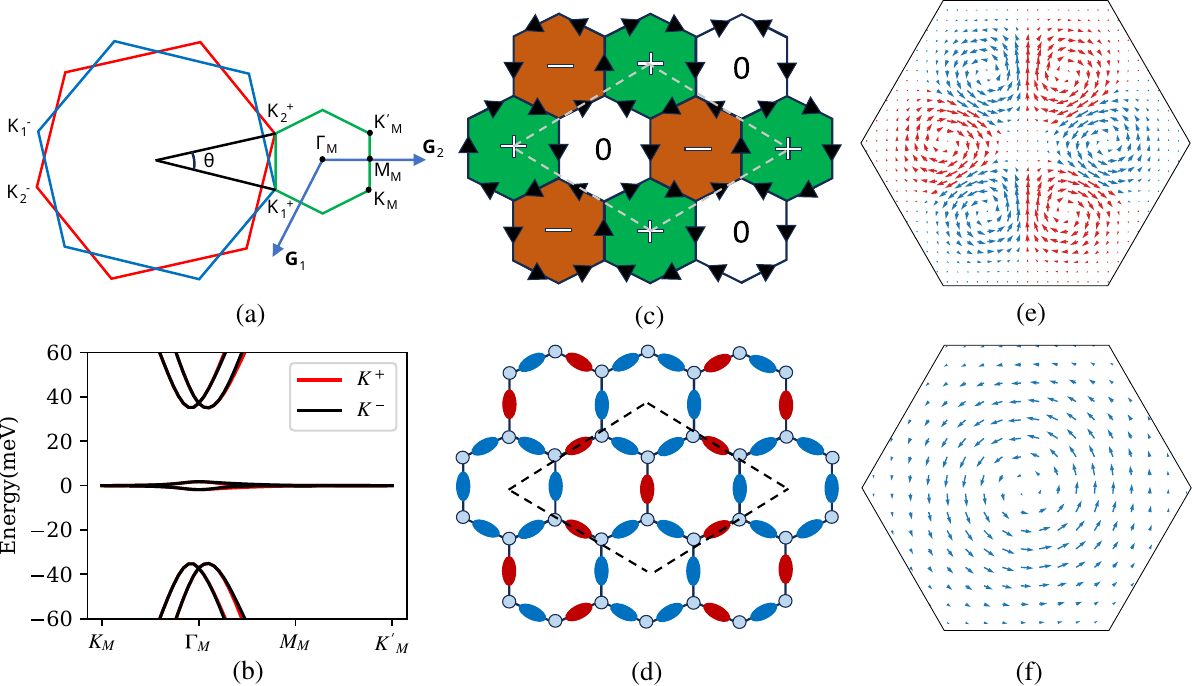}    
\end{center}

\caption{(a) Red and blue hexagons are the Brillouin zones of the two graphene monolayers. The green hexagon denotes the mBZ. $\textbf{G}_1$ and $\textbf{G}_2$ are the \moire reciprocal lattice vectors. (b) The band structure of twisted bilayer graphene at $\theta=1.9^\circ$. (c) The schematic of the KIVC phase. The + (-) sign represents the clockwise (anticlockwise) current around the graphene unit cell. The 0 represents net zero current. The dashed lines represents the $\sqrt{3}\times\sqrt{3}$ order due to the tripling of the unit cell. (d) The schematic of the TIVC phase. The different color shades of the graphene bonds represent different values of the bond coherence. The dashed lines represent the $\sqrt{3}\times\sqrt{3}$ order due to the tripling of the unit cell. (e) The current pattern of the VP phase within a \moire unit cell. (f) The current pattern of the QAH phase within a \moire unit cell.}
\label{Schematics_phase}
\end{figure*}

Twisted bilayer graphene is formed by twisting the top and bottom layers of an aligned bilayer graphene by $+\theta/2$ and $-\theta/2$, respectively.  In addition to their momenta, the Bloch states in each graphene layer can be described by a sublattice ($\sigma=A,B$) and a spin ($s$) quantum number. We will consider the low-energy dispersion of the graphene layers around the band-touching points $K$ and $K’$, which adds a valley quantum number $\tau=\pm 1$ for $K$ and $K’$ valleys. We will later also use Pauli matrices $\hat{\tau}_{x/y/z}$, $\hat{\sigma}_{x/y/z}$ and $\hat{s}_{x/y/z}$, which will operate in the corresponding two-dimensional spaces. The interlayer hopping will tile the original graphene Brillouin zone with the mBZ with primitive reciprocal vectors $\textbf{G}_1=(-\frac{2\pi}{\sqrt{3}L_M},-\frac{2\pi}{L_M})$ and $\textbf{G}_2=(\frac{4\pi}{\sqrt{3}L_M},0)$, as shown in Fig.~\ref{Schematics_phase}(a), where $L_M$ is the \moire lattice constant. They are related to the momentum difference between the Dirac points in the two misaligned layers. We introduce the index $X$ which denotes the combined indices for sublattice and layer, $\{A1,B1,A2,B2\}$. In this layer-sublattice ($X$) basis the noninteracting system is described by the following continuum Hamiltonian \cite{Bistritzer_PNAS_2011, Santos_PRL_2007, Santos_PRB_2012, Koshino_PRX_2018}

\begin{equation}
H_{0}^\tau=\begin{pmatrix}
-\hbar v_F(\hat{\textbf{k}}-\textbf{K}_1^\tau).\boldsymbol{\sigma}^\tau & U_\tau\\
U^\dagger_\tau & -\hbar v_F(\hat{\textbf{k}}-\textbf{K}_2^\tau).\boldsymbol{\sigma}^\tau
\end{pmatrix}
\end{equation}

where $\hat{\textbf{k}}=-i\partial_\textbf{r}$ and $\boldsymbol{\sigma}^\tau=(\tau \hat{\sigma}_x, \hat{\sigma}_y)$. $\textbf{K}_1^\tau$ and $\textbf{K}_2^\tau$ denote the Dirac points of the two rotated layers as shown in Fig.~\ref{Schematics_phase}(a). In the rest of the paper, $K$ \& $K'$ and $K^{+}$ \& $K^{-}$ have been used interchangeably. In this work, the monolayer graphene lattice constant is taken as $a_0$=0.246 nm. We take $\hbar v_F$=5817.9 meV$\cdot$\AA~ throughout the paper \cite{Fuchun_PRB_2020}.

The interlayer tunnelling is given by 
\begin{multline}
U_\tau=\begin{pmatrix}
w_{AA} & w_{AB}\\
w_{AB} & w_{AA}
\end{pmatrix} 
+
\begin{pmatrix}
w_{AA} & w_{AB}e^{-\tau \frac{2\pi i}{3}}\\
w_{AB}e^{\tau \frac{2\pi i}{3}} & w_{AA}
\end{pmatrix}
e^{-i\tau \textbf{G}_1.\textbf{r}}\\
+
\begin{pmatrix}
w_{AA} & w_{AB}e^{\tau \frac{2\pi i}{3}}\\
w_{AB}e^{-\tau \frac{2\pi i}{3}} & w_{AA}
\end{pmatrix}
e^{-i\tau (\textbf{G}_1+\textbf{G}_2).\textbf{r}}
\end{multline}

The intersublattice ($w_{AB}$) and intrasublattice ($w_{AA}$) interlayer couplings are chosen to be 110 meV and 80 meV, respectively \cite{Fuchun_PRB_2020}. The Bloch states of the noninteracting system are obtained by diagonalizing $H^\tau_{0}$

\begin{equation}
\psi^X_{\tau \eta \textbf{k}}(\textbf{r})=\sum_{\textbf{G}}u_{\tau,\eta;\textbf{G},X}(\textbf{k})e^{i(\textbf{k}+\textbf{G}).\textbf{r}}
\end{equation}

with eigenenergy $\epsilon_{\tau\eta}(\textbf{k})$. Here $\eta$ is the index of the resulting moire minibands in the system.

With our chosen parameters, the magic twist angle for MATBG is $1.09^\circ$ and we will fix $\theta$ to this value in this paper. The bandwidth of the low energy flat bands at the magic angle is $\sim 2$ meV with a bandgap to higher bands $\sim 35$ meV as shown in Fig.~\ref{Schematics_phase}(b). Hence, for each spin and valley component, we focus only on the two low-energy bands ($\eta=1$ for conduction and $\eta=-1$ for valence band). Thus for each momenta in the mBZ, we have 8 degrees of freedom. As before, we will define Pauli matrices $\hat{\eta}_{x/y/z}$ which operate in the two-dimensional space of conduction-valence bands.
 
 We note that the sublattice and band degrees of freedom provide alternate descriptions of the same low energy physics and one can easily go from one basis to the other through  \cite{Bernevig_PRB_2021}
 \begin{eqnarray}
 \begin{aligned}
      f^{\dagger}_{KA(B)s}(\textbf{k})&=&\frac{1}{\sqrt{2}}(d^{\dagger}_{KCs}(\textbf{k})\pm i d^{\dagger}_{KVs}(\textbf{k}))\\ 
      f^{\dagger}_{K'A(B)s}(\textbf{k})&=&\frac{1}{\sqrt{2}}(d^{\dagger}_{K'Cs}(\textbf{k})\mp i d^{\dagger}_{K'Vs}(\textbf{k})).
\end{aligned}      
 \end{eqnarray}    

where $C$ and $V$ denote conduction and valence bands, respectively. $f^\dagger (f)$ denotes creation (annihilation) operator in the valley-sublattice-spin basis. Similarly, $d^\dagger (d)$ denotes creation (annihilation) operator in the valley-band-spin basis.

\subsection{Projected Coulomb interaction} 
The quenching of the bandwidth at the magic angle makes the electron-electron (e-e) interactions the key element determining the low-energy physics of the system. As discussed in the Introduction, e-e can induce many broken-symmetry phases depending on the carrier filling. To study these symmetry-broken phases, we use HF theory, which has emerged as a successful tool to study these symmetry-broken phases \cite{Bultinck_PRX_2020, Wagner_PRL_2022, Bernevig_PRB_2021, Fuchun_PRB_2020, Xie_PRL_2020, Cea_PRB_2020, Liu_PRB_2021}.  

In the actual experimental setup, the MATBG sample is either sandwiched between two metallic gates (dual-gate configuration) or placed on top of a metallic gate (single-gate configuration). hBN is used as a spacer between the sample and the gates. The thickness ($d$) of the hBN spacer is termed the gate distance. The presence of metallic gates alters the bare Coulomb interaction through the formation of image charges. In this work, we consider the scenario of a single-gate setup. In the single-gate setup, the gate-screened Coulomb potential  becomes

\begin{equation}
V(\textbf{q})=\frac{e^2}{2\epsilon\epsilon_0q}(1-e^{-2qd})
\label{Coulomb_pot}
\end{equation}

where $\epsilon=7$ is the dielectric constant of the hBN substrate. $d$ can be varied to tune the strength of the Coulomb interaction \cite{Goodwin_PRB_2020}. Experiments reveal that for $d$ smaller than the \moire length scale leads to higher screening of e-e interaction, resulting in the suppression or disappearence of the correlated insulating states \cite{Stepanov_Nature_2020}.

For the HF study, we follow the formalism of ref.\cite{Fuchun_PRB_2020}. Only the eight low-energy flat bands are considered. The gate-screened Coulomb interaction of Eq.~\eqref{Coulomb_pot} is projected only on these eight bands. The projection can be justified by noting that the high-energy bands are well separated from the flat bands for the realistic ratio $w_{AA}/w_{AB} \sim 0.8$ (Fig.~\ref{Schematics_phase}(b)).  The projected Hamiltonian has the following form

\begin{equation}
    \begin{split}
		H &=\sum_{\tau, \eta, s, \textbf{k}} \epsilon_{\tau\eta}(\textbf{k})\,d^{\dagger}_{\tau \eta s}(\textbf{k})\,d_{\tau \eta s}(\textbf{k})\\
		&\quad +\frac{1}{2\Omega}	\sum_{\eta_i}\,\sum_{\textbf{k},\textbf{k}',\textbf{q}}\,\sum_{s,s',\tau, \tau'}\,V_{\eta_1 \eta_2 \eta_3 \eta_4}^{\tau \tau' \tau' \tau}(\textbf{k},\textbf{k}',\textbf{q})\\
		&\quad d^{\dagger}_{\tau \eta_1 s}(\textbf{k}-\textbf{q})\,d^{\dagger}_{\tau' \eta_2 s'}(\textbf{k}'+\textbf{q})\,d_{\tau' \eta_3 s'}(\textbf{k}')\,d_{\tau \eta_4 s}(\textbf{k})
	\end{split}
\label{Ham_tot}	
\end{equation}	

where $\eta_i \in$\{V,C\}, $\textbf{k},\textbf{k}',\textbf{q} \in$ mBZ. $\Omega$ is the area of a \moire unit cell. We neglect interaction processes that involve intervalley scattering, as it involves a large transfer of momentum $\textbf{q}\sim(K-K’)$ is approximately the size of the graphene Brillouin zone, and the Coulomb matrix elements would be exponentially suppressed in that case. The projected Coulomb matrix element $V_{\eta_1 \eta_2 \eta_3 \eta_4}^{\tau \tau' \tau' \tau}(\textbf{k},\textbf{k}',\textbf{q})$ is given by the following expression

\begin{multline}
	V_{\eta_1 \eta_2 \eta_3 \eta_4}^{\tau\tau'\tau'\tau}(\textbf{k},\textbf{k}',\textbf{q})=\sum_{\textbf{G}}\,V(\textbf{q}+\textbf{G})\\\lambda_{\eta_1,\eta_4;\tau}(\textbf{k}-\textbf{q},\textbf{k}+\textbf{G})\,\lambda_{\eta_3,\eta_2;\tau'
}^*(\textbf{k}',\textbf{k}'+\textbf{q}+\textbf{G}).
\end{multline}

where the form factor $\lambda_{\eta_1,\eta_2;\tau}(\textbf{k},\textbf{k}'+\textbf{G})$ is the overlap of the noninteracting wavefunctions

\begin{equation}
\lambda_{\eta_1,\eta_2;\tau}(\textbf{k},\textbf{k}'+\textbf{G})=\sum_{\textbf{G}',X}u^{*}_{\tau,\eta_1;\textbf{G}',X}(\textbf{k})\,u^{*}_{\tau,\eta_2;\textbf{G}+\textbf{G}',X}(\textbf{k}')
\end{equation}

where $\textbf{G}=n_1\textbf{G}_1+n_2\textbf{G}_2$. In our calculation, we have used $n_1,n_2 \in$ [-5,5] with the restriction that each \textbf{G} needs to have its $C_3$ partners in order to preserve the $C_3$ symmetry. Our interacting model has a $SU(2)$ spin rotation invariance together with a $U_V(1)$ valley charge conservation and a $U_C(1)$ total charge conservation. It is also invariant under a $C_3$ rotation, a combination of inversion and time reversal $C_2T$, and the spinless time reversal operators $T=\tau_xK$ and $T'=\tau_yK$, where $K$ is the complex conjugation \cite{Bernevig_PRB_2021, Po_PRX_2018, Bultinck_PRX_2020}. The time reversal operators also change the momentum $\textbf{k}$ in mBZ to $-\textbf{k}$. The two-fold rotation $C_{2y}$ along the $y-$axis is also a symmetry, which exchanges the two layers and the two valleys $K$ and $K'$ \cite{Bernevig_PRB_2021}.

\subsection{Hartree-Fock mean-field theory}

We decouple the interaction term in Eq.~\eqref{Ham_tot} using Hartree-Fock theory to write a mean-field Hamiltonian

\begin{multline}
    H_{MF} = \sum_\textbf{k} \sum_{\tau,\eta,s} \sum_{\tau’,\eta', s’} d^\dagger_{\tau\eta s}(\textbf{k})  [  [ \epsilon_{\tau\eta}(\textbf{k}) -\mu ]\delta_{\tau\tau’} \delta_{\eta\eta’}\delta_{ss’} \\
    +h_{\tau s}^{\eta\eta’} (\textbf{k}) \delta_{\tau\tau’} \delta_{ss’}  +\chi^{\eta\tau s;\eta’\tau’ s’} (\textbf{k}) ] d_{\tau’\eta’s'}(\textbf{k}) 
    \label{Ham_meanfield}
\end{multline}

where the Hartree-shift, $h^{\eta\eta'}_{\tau s}$, and the Fock-shift $\chi^{\eta\tau s;\eta’\tau’ s’}$, are given by the following expressions

\begin{multline*}
h^{\eta\eta'}_{\tau s}=\frac{1}{\Omega}\sum_{\eta'',\eta''',\textbf{k}', s',\tau'}V^{\tau \tau' \tau' \tau}_{\eta \eta'' \eta''' \eta'}(\textbf{k},\textbf{k}',0)\\
\left(\left \langle d^{\dagger}_{\tau' \eta'' s'}(\textbf{k}')d_{\tau' \eta''' s'}(\textbf{k}')\right\rangle - \left \langle d^{\dagger}_{\tau' \eta'' s'}(\textbf{k}')d_{\tau' \eta''' s'}(\textbf{k}')\right\rangle_0\right)
\end{multline*}

\begin{multline}
\chi^{\eta\tau s;\eta’\tau’ s’}=-\frac{1}{\Omega}\sum_{\eta'',\eta''',\textbf{k}'}V^{\tau' \tau \tau \tau'}_{\eta'' \eta \eta''' \eta'}(\textbf{k},\textbf{k}',\textbf{k}-\textbf{k}')\\
\left(\left \langle d^{\dagger}_{\tau\eta'' s}(\textbf{k}')d_{\tau' \eta''' s'}(\textbf{k}')\right\rangle - \left \langle d^{\dagger}_{\tau' \eta'' s'}(\textbf{k}')d_{\tau' \eta''' s'}(\textbf{k}')\right\rangle_0\right)
\label{HF_shift}
\end{multline}

We solve Eq.~\eqref{Ham_meanfield} self-consistently to obtain the ground state of the system. In this work, we have only looked for phases that preserve the translational symmetry of the system.

Note that the noninteracting model already includes the effects of Coulomb interaction to some extent, as the values of $w_{AA}$ and $w_{AB}$ are determined either from experiments or from density functional theory-based calculations, which include Coulomb interactions. Hence, to avoid double counting of the interaction, a reference density matrix is subtracted from the calculated density matrix in both the Hartree and Fock shift terms in Eq.~\eqref{HF_shift}. In the literature various different subtraction schemes have been used: (i) decoupled graphene scheme \cite{Bultinck_PRX_2020, Xie_PRL_2020} (the reference density matrix is calculated at the charge neutrality point of the two decoupled graphene layers), (ii) CNP scheme \cite{Hejazi_PRR_2021} (the reference density matrix is calculated at the charge neutrality point of the noninteracting MATBG system), (iii) average scheme \cite{Kang_PRL_2021, Bernevig_PRB_2021} (the reference density matrix is calculated assuming half filling of every states of the eight flat bands), and (iv) variable scheme \cite{Fuchun_PRB_2020} (for a given filling the reference density is calculated at the same filling of the noninteracting MATBG system).  In the main paper, we report the results of the CNP subtraction scheme. For the results of the other subtraction schemes refer to the supplementary material \cite{Supple}.
 
Another important aspect of the calculation is the gauge fixing of the noninteracting wavefunctions. It is necessary to get the correct phases of the order parameters of the various broken symmetry states. The phases of the conduction- and valence-band wavefunctions  in the $K$ and $K'$ valleys of the noninteracting system are fixed using various symmetries of the system \cite{Bernevig_PRB_2021}. The details can be found in the supplementary material \cite{Supple}.

\section{Phase diagram}

\begin{table}
   \caption{Order parameters of various phases} 
   \label{table}
   \small
   \centering
   \begin{tabular}{lccr}
   \toprule\toprule
   \textbf{Phase} & \textbf{Sublattice basis} & \textbf{Band basis} \\ 
   \midrule
   KIVC & $\tau_{x,y} \sigma_y s_0$  & $\tau_{y,x} \eta_y s_0$\\
   TIVC (bond) & $\tau_{x,y} \sigma_{x} s_0$  & $\tau_{x,y} \eta_{0} s_0$\\
   TIVC (site) & $\tau_{x,y} \sigma_{0} s_0$  & $\tau_{x,y} \eta_{z} s_0$\\
   VP & $\tau_z\sigma_0 s_0$ & $\tau_z\eta_0 s_0$ \\
   QAH & $\tau_z\sigma_z s_0$  & $\tau_z(\eta_0-\eta_y) s_0$\\
   \bottomrule
   \end{tabular}
   \label{table_1}
\end{table}

\begin{figure*}[t]
\begin{center}
\includegraphics[scale=0.8]{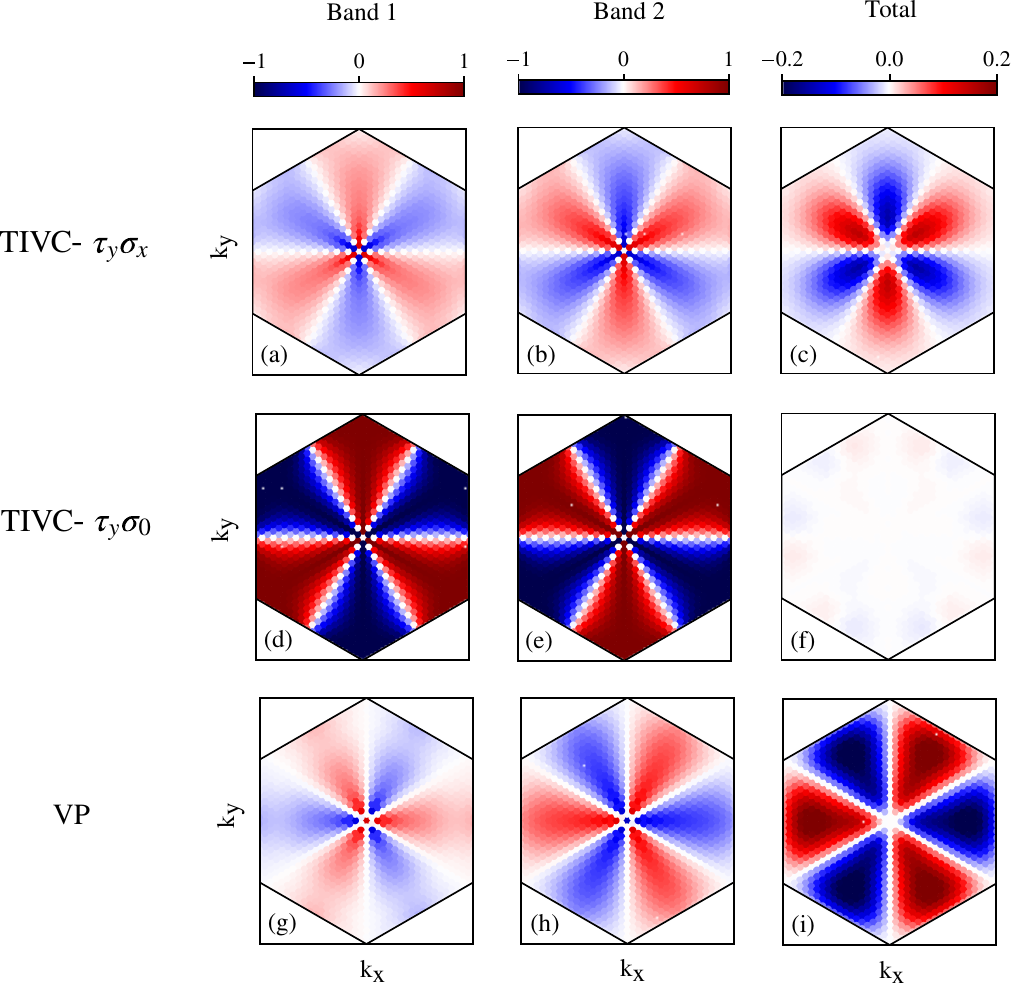}    
\end{center}
\caption{Order parameters of the ground state in the mBZ at $\nu=-2.0$ at $d$ = 100 nm. (a)--(c) Show the TIVC-$\tau_y\sigma_x$ order for band 1, band 2, and the two bands combined, respectively. See the main text for the definitions of the bands. (d)--(f) Show the TIVC-$\tau_y\sigma_0$ order for band 1, band 2, and the two bands combined, respectively. (g)--(i) Show the VP order for band 1, band 2, and the two bands combined, respectively. The colorbar above each column applies to all the figures in the column.}
\label{fwave_-2.0}
\end{figure*}

In this section, we will describe the symmetry-broken phases and phase transitions of MATBG when its carrier density is being tuned between $\nu=-2$ and $\nu=-3$. We will study these phases as a function of the gate distance $d$, which will be a proxy for the effective strength of electronic interactions. We will present the phase diagram in the $\nu-d$ plane at $T=0$. Here, we would like to mention that Wagner et al. \cite{Wagner_PRL_2022} reported a study of the ground-state phases between $\nu=-2$ and $\nu=-3$ as a function of strain at a fixed $d$. 

In the absence of strain, the HF theory produces spin-polarized SU(2) broken states as the low-energy solutions. The states are distinguished in terms of valley and sublattice couplings with order parameters $\Delta(\textbf{k}) \sim \langle f^\dagger_{\tau\sigma s}(\textbf{k}) \vec{\tau}_{\tau\tau'} \vec{\sigma}_{\sigma\sigma'}f_{\tau'\sigma' s}(\textbf{k}) \rangle$. Here, we will not consider incommensurate orders with a finite net momentum, which are stabilized only in presence of strain \cite{Kwan_PRX_2021}.

Let us first consider the valley degree of freedom. A class of symmetry-broken phases are the inter-valley coherent (IVC) states, which couple the two valleys. The interacting bands are a linear combination of states in the two valleys, and they break the $U_V(1)$ valley charge conservation. Since the graphene $K$ and $K'$ points are not related by a Brillouin zone vector, the IVC states lead to a tripling of the graphene unit cell, creating a $\sqrt{3} \times \sqrt{3}$ Kekulé modulation on the graphene lattice spacing scale. The IVC states can be further classified on the basis of how the order parameter couples the sublattice degrees of freedom. For the Kramers IVC (KIVC) state, the ordering creates alternating patterns of circulating currents on the graphene bonds which are modulated with the Kekulé pattern \cite{Bultinck_PRX_2020}. (see Fig.~\ref{Schematics_phase}(c) for a schematic representation). This state is thus equivalent to an orbital magnetization density wave in the system. The order parameter $\Delta \sim \tau_{x,y} \sigma_y$ breaks the spinless time-reversal symmetry generated by $T =\tau_x K$, but is invariant under the modified time reversal generated by $T'=\tau_y K$. For the KIVC state, it is also useful to look at the order parameter in the non-interacting band basis instead of the sublattice basis. Here $\Delta \sim\tau_{y,x} \eta_y$, i.e. the KIVC state mixes the valence and the conduction band, leading to the formation of intervalley excitons.

The other prominent IVC phase is the time-reversal symmetric IVC (TIVC) with an order parameter $\Delta\sim \tau_{x,y}\sigma_x$, which generates a bond density wave state with Kekulé modulation, schematically shown in Fig.~\ref{Schematics_phase}(d). There is a modified version of TIVC with $\Delta\sim \tau_{x,y}\sigma_0$, which generates a charge density wave with Kekulé modulation on the graphene lattice scale. The TIVC phase breaks the $T'$ symmetry while preserving the $T$ symmetry in the system. In the non-interacting band basis, the TIVC order mixes the valence (or conduction) bands of different valleys amongst themselves, leading to a standard density wave picture.

Instead of the IVC phases, we can also consider the valley-polarized (VP) phase, where electrons preferentially occupy states in either the $K$ or the $K'$ valley, with an order parameter $\Delta \sim \tau_z$. A uniform VP phase breaks both $T$ and $T'$ symmetry, but satisfies $C_2T$ and $C_2T'$. The schematic in Fig.~\ref{Schematics_phase}(e) shows the real-space current pattern in the VP phase. There are six alternating circulating current patterns in the \moire unit cell, and the phase is dubbed the \moire orbital antiferromagnet \cite{Liu_PRB_2021}.

The VP state can further break sublattice symmetry, so that electrons preferentially occupy states with either $A$ or $B$ sublattice (on top of preferentially occupying $K$ or $K'$ valley). This is the quantum anomalous Hall (QAH) phase, where a single current loop is generated in the moire unit cell \cite{Liu_PRB_2021} (see Fig.~\ref{Schematics_phase}(f) for a schematic representation). The QAH phase supports a finite orbital ferromagnetism in the system \cite{Liu_PRB_2021, Bultinck_PRL_2020}. In the non-interacting band basis, the sublattice polarized state mixes the valence and conduction bands, while the ordinary VP state is diagonal in the band basis. The QAH phase breaks $C_2T (T')$, which leads to the anomalous Hall response.

The above discussion of the symmetry of the states assumed a $s$-wave order parameter, i.e. it assumed $\Delta(\textbf{k})$ is almost independent of \textbf{k} (or at the least does not break additional symmetries). However, there is no a-priori reason to believe that the order parameters do not have any non-trivial form factors. For the effective triangular \moire lattice, the most natural non-trivial form factor is the $f_{x/y}$-wave given by $\Delta(\textbf{k}) = \sin(k_xa)-2\sin(\frac{k_xa}{2})\cos(\frac{\sqrt{3}k_ya}{2})/\sin(\sqrt{3}k_ya)-2\sin(\frac{\sqrt{3}k_ya}{2})\cos(\frac{3k_xa}{2})$. The $f$-wave form factor acquires a minus sign on a $\pi/3$ rotation and forms a projective representation for the triangular lattice.

If we assume a $f$-wave form factor, the symmetries of the different phases considered above changes. The key difference is that changing the momentum \textbf{k} to -\textbf{k} in the mBZ acquires an additional minus sign due to the form factor (Note that this will be true for all odd parity form factors). For $f$-wave orders, the KIVC phase breaks $T'$ symmetry and is invariant under $T$, while the TIVC states are invariant under $T'$, and they break the $T$ invariance. The VP state with $f$-wave form factor will be invariant under both $T$ and $T'$.

\begin{figure*}[t]
\begin{center}
\includegraphics[scale=0.8]{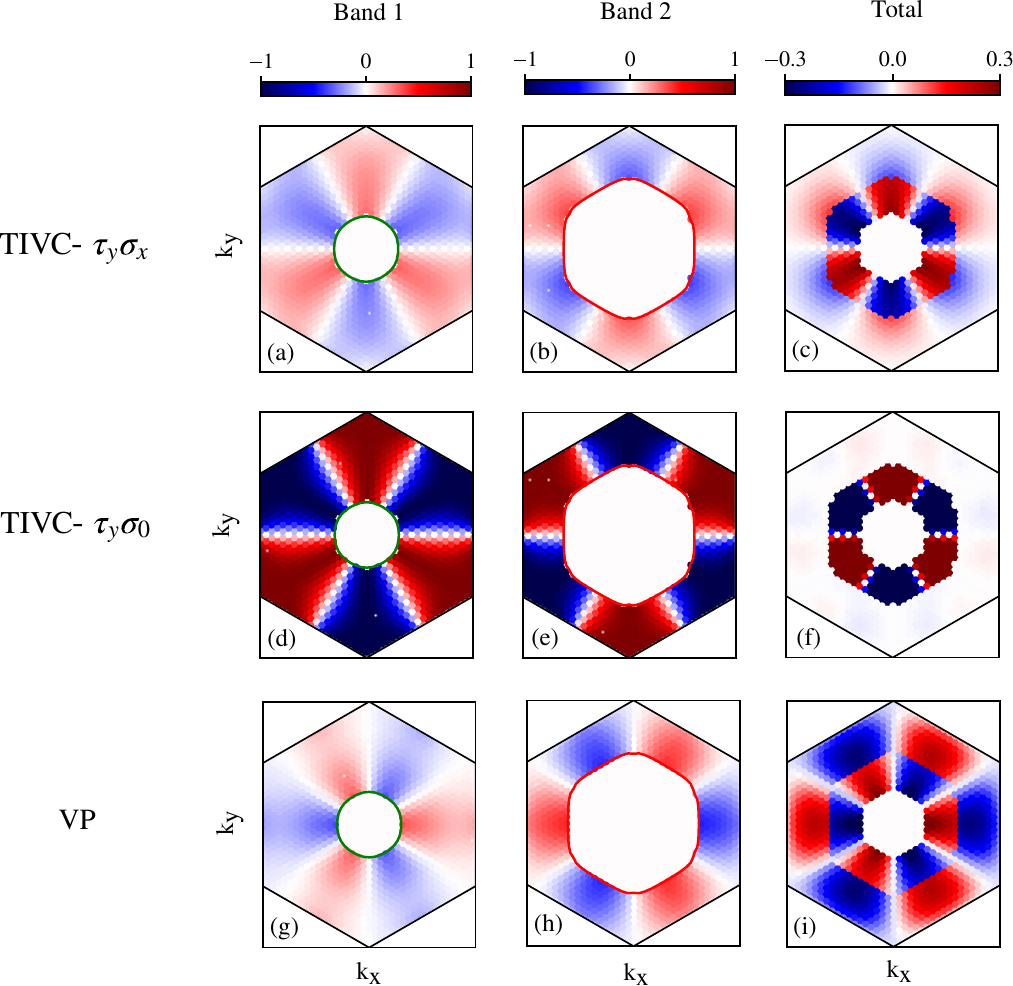}    
\end{center}
\caption{Order parameters of the ground state in the mBZ at $\nu=-2.40$ at $d$ = 100 nm. (a)--(c) Show the TIVC-$\tau_y\sigma_x$ order for band 1, band 2, and the two bands combined, respectively. See the main text for the definitions of the bands. The green line in (a) and the red line in (b) denote the Fermi contours of the respective bands.  (d)--(f) Show the TIVC-$\tau_y\sigma_0$ order for band 1, band 2, and the two bands combined, respectively. The green line in (d) and the red line in (e) denote the Fermi contours of the respective bands. (g)--(i) Show the VP order for band 1, band 2, and the two bands combined, respectively. The green line in (g) and the red line in (h) denote the Fermi contours of the respective bands. The colorbar above each column applies to all the figures in the column.}
\label{fwave_-2.40}
\end{figure*}

\subsection{Insulating state at $\boldsymbol\nu$=-2}

The insulating state at $\nu=-2$ has been studied using HF theory \cite{Fuchun_PRB_2020, Xie_PRL_2020, Liu_PRB_2021, Kwan_PRX_2021, Zhang_PRL_2022, Kwan_PRB_2024}, density matrix renormalization group \cite{Soejima_PRB_2020}, exact diagonalization \cite{Potasz_PRL_2021, Xie_PRB_2021}, quantum Monte Carlo \cite{Xiao_PRR_2025}, and dynamical mean-field theory \cite{Rai_PRX_2024} techniques. In the absence of strain, it was found that the spin-polarized KIVC state formed the ground state of the system at this commensurate filling. Our HF calculations also reproduce this. The KIVC correlator $\langle \psi^\dagger_\textbf{k} \tau_y \sigma_y \psi_\textbf{k}\rangle$ can be plotted as a function of $\textbf{k}$ and shows a $s$-wave (uniform) form factor. At $\nu=-2$, there are two occupied bands in the HF theory. One can also consider the contribution of each of these HF bands (not to be confused with the two non-interacting bands) to the order. We find that both these bands contribute almost equally, i.e., the KIVC order is independent of the momentum and band indices.

While the KIVC order has been discussed extensively in the literature, the HF theory also predicts subdominant TIVC and VP correlations, which have not been discussed before to the best of our knowledge.  Let us first focus on the TIVC correlations given by $\langle \psi^\dagger_\textbf{k} \tau_y \sigma_x \psi_\textbf{k}\rangle$. We can further divide this correlation into contributions from the two occupied HF bands. The contribution from each of these bands and their sum is plotted as a function of $\textbf{k}$ in Figs.~\ref{fwave_-2.0}(a)-(c). Here, band1 and band2 represent the lowest and second-lowest filled HF bands, respectively (see Fig.~\ref{Bandstructure}(a)). We see that each band hosts a TIVC order with a $f$-wave form factor with nodes along the $x$-axis and the two other symmetry related directions at $60^\circ$ and $120^\circ$. However, the contributions from the two bands are out of phase, i.e., for $\textbf{k}$ points where the contribution of the first band is positive, that of the second band is negative. However, the band correlations are not equal and opposite; i.e., when summed over, we are left with a sizeable TIVC order with a $f$-wave form factor ( for an effective triangular \moire lattice, $f$-wave gives a projective representation, while $p$ and $d$ wave form factors break additional lattice symmetries). To get a quantitative estimate, we note that this correlation is $\sim 0.2$ (see Fig.~\ref{fwave_-2.0}(c)), whereas the $s$-wave KIVC order is $\sim 0.9$. Thus, the $f$-wave TIVC is subdominant but not negligible in this state. It is important to note that unlike a $s$-wave TIVC order, which preserves $T$ symmetry but violates the $T'$ symmetry, the $f$-wave TIVC satisfies $T'$ symmetry and violates the $T$ symmetry (just like the $s$-wave KIVC order).

One can also consider another TIVC order, given by $\langle \psi^\dagger_\textbf{k} \tau_y \sigma_0 \psi_\textbf{k}\rangle$. This is shown in Figs.~\ref{fwave_-2.0}(d)-(f). In this case, the individual HF bands host a finite TIVC order with a $f$-wave symmetry, but they are equal and opposite (i.e., perfectly phase shifted). Hence the total contribution of this TIVC order vanishes, as seen in Fig.~\ref{fwave_-2.0}(f).

The HF bands also host a valley polarization with a $f$-wave form factor (with nodes along $y$-axis and symmetry-rotated directions). This is shown in Fig.~\ref{fwave_-2.0}(g)-(i); e.g., the first band has predominantly $K$ valley states in the angles between $\pm \pi/6$, while the second band has predominantly $K'$ valley states in this region. This polarization flips after every $60^\circ$, leading to the $f$-wave form factor. The net valley polarization of the momentum states also has a $f$-wave form with a peak value of $\sim 0.2$. as seen in Fig.~\ref{fwave_-2.0}(i). A $f$-wave valley polarization also preserves $T'$ symmetry. 

We thus see that while the $\nu=-2$ state is predominantly a KIVC state, other $T'$ invariant orders with $f$-wave form factors (TIVC and VP) are also present in the system. In fact, the contribution of the individual occupied HF bands to these orders is often ${\cal O}(1)$, but they sum up with opposite signs, leaving a subdominant role for these secondary orders.

\subsection{Metallic phase near $\boldsymbol\nu$=-2}

\begin{figure*}[t]
\begin{center}
\includegraphics[scale=.85]{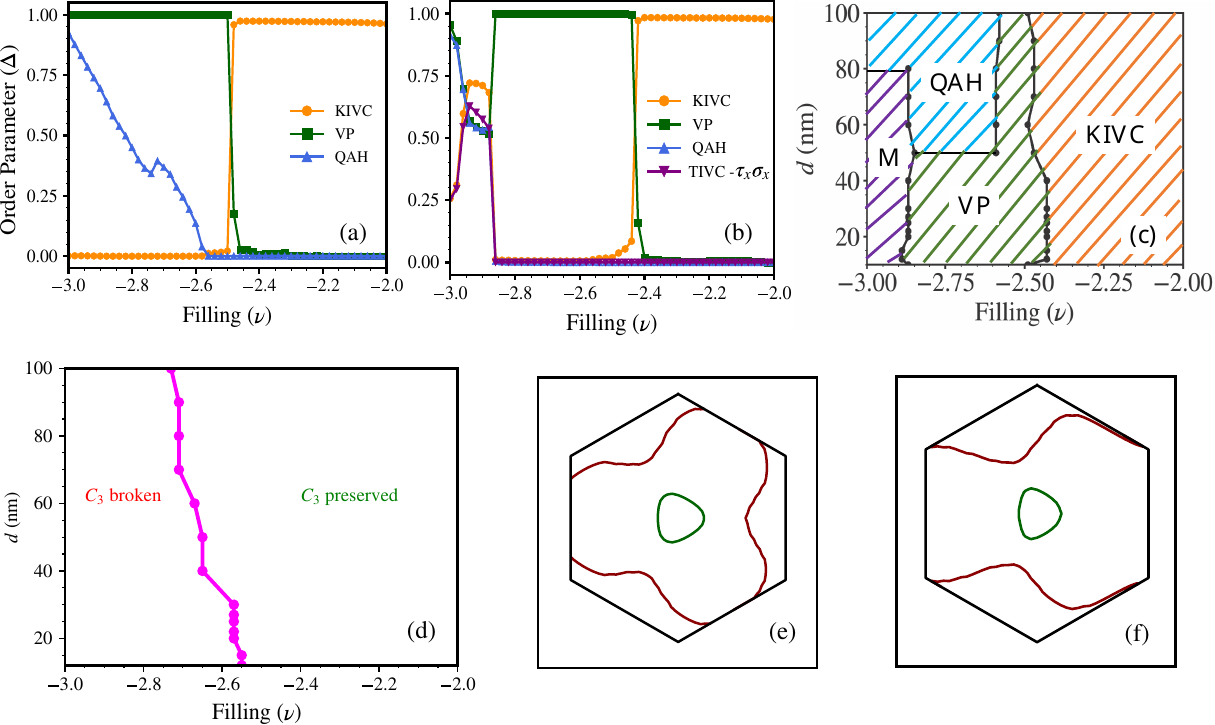}   
\end{center}
\caption{(a) Plot of $s-$wave component of various order parameters as a function of $\nu$ at $d=100$ nm. (b) Same as (a) at $d=20$ nm. (c) The phase diagram as function of $\nu$ and $d$. Here we show only the dominant phase at a given $\nu$ and $d$. (d) Phase-space of $C_3$ symmetry. The solid line separates the two regions where $C_3$ is satisfied and where it breaks down. (e) The Fermi contours at $d=100$ nm at $\nu=-2.72$ (QAH phase). (f) The Fermi contours at $d=100$ nm at $\nu=-2.74$ (QAH phase).}
\label{Phase_diagram_dominant_order}
\end{figure*}

The KIVC state at $\nu=-2$ is robust to further hole doping of the system. In fact, the average KIVC order parameter $\frac{1}{N_e}\sum_\textbf{k} \langle \psi^\dagger_\textbf{k} \tau_y \sigma_y \psi_\textbf{k} \rangle$ ($N_e\equiv$ no of electrons) remains almost independent of the doping, till the density reaches $\nu =-2.48$, when it sharply drops to zero, and the system undergoes a first-order phase transition to a VP state (see Fig.~\ref{Phase_diagram_dominant_order}(a)). We now focus our attention on how the subleading $f$-wave orders change with doping in this metallic KIVC phase. 

\begin{figure}
\vspace{2ex}
 \begin{center}
 \includegraphics[scale=1.1]{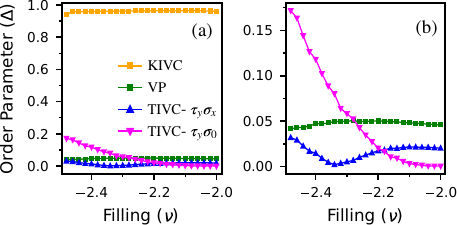}
 \end{center}
 \vspace{-3ex}
 \caption{Order parameters of the ground state near $\nu=-2$. (a) Plot of $s-$wave KIVC ($\tau_y\sigma_y$), $f_y-$wave VP, $f_x-$wave TIVC-$\tau_y\sigma_x$, and $f_x-$wave TIVC-$\tau_y\sigma_0$ order parameters as function of $\nu$. (b) Magnified plot of VP, TIVC-$\tau_y\sigma_x$, and TIVC-$\tau_y\sigma_0$ order parameters from (a).}
  \label{Orderpara_IVC_phase}
\end{figure}

As the system is doped, hole pockets around the $\Gamma_M$ point appear in both HF bands. However, the size of these pockets in the two bands are different. Representative Fermi surfaces in the two bands at $\nu =-2.4$ are shown with red and green lines in the mBZ  in Fig.~\ref{fwave_-2.40}. Let us first focus on the TIVC order represented by $\tau_y\sigma_0$. In the insulating phase, the contributions of the two HF bands were large, but equal and opposite, leading to a vanishing value for this order when the band contributions were summed. As we dope the system, this balance is violated in the region between the two Fermi surfaces, leading to an increase in this $f$-wave order. The order increases with doping as the difference in the size of the Fermi surfaces grows with doping. The contributions at each momenta are plotted in Figs.~\ref{fwave_-2.40}(d)-(f). We see that the contribution from inside the hole pockets vanish as these states are not occupied. The $f$-wave order becomes appreciable in the region between the two Fermi surfaces. One can construct an average $f$-wave TIVC $\tau_y\sigma_0$ order parameter through $\frac{1}{N_e}\sum_\textbf{k}  \tilde{\chi}_\textbf{k} \langle \psi^\dagger_\textbf{k} \tau_y \sigma_0 \psi_\textbf{k} \rangle$, where $\tilde{\chi}_\textbf{k}$ is the normalized form factor  $\tilde{\chi}_\textbf{k}=\chi_\textbf{k}/\sqrt{N_\textbf{k}^{-1}(\sum_{\textbf{k}'}|\chi_{\textbf{k}'}|^2})$ and $\chi_\textbf{k} = \sin(k_xa)-2\sin(\frac{k_xa}{2})\cos(\frac{\sqrt{3}k_ya}{2})$ is the $f_x$-wave form factor ($N_\textbf{k}\equiv$ no of \textbf{k} points in mBZ). This quantity is plotted in Fig.~\ref{Orderpara_IVC_phase}(a) and (b) as a function of doping. It is clear that this order increases with doping the metallic state and reaches about 18\% of the dominant KIVC order parameter around $\nu =-2.4$.

The situation is reversed in the case of the $\tau_y\sigma_x$ TIVC order and the valley polarization, which had a finite net $f$-wave component in the insulating case. In these cases, the region between the Fermi surfaces contribute with a sign opposite to that of the net $f$-wave order at $\nu=-2$. This is clearly seen in Figs.~\ref{fwave_-2.40}(a)-(c) for the TIVC $\tau_y\sigma_x$ order and in Figs.~\ref{fwave_-2.40}(g)-(i) for the valley polarization. As a result, these $f$-wave orders do not change substantially with doping. The average $f$-wave order parameter, for the TIVC order is given by $\frac{1}{N_e}\sum_\textbf{k} \tilde{\phi}_\textbf{k} \langle \psi^\dagger_\textbf{k} \tau_y \sigma_x \psi_\textbf{k} \rangle$, where $\tilde{\phi}_\textbf{k}$ is the normalized form factor $\tilde{\phi}_\textbf{k}=\phi_\textbf{k}/\sqrt{N_\textbf{k}^{-1}(\sum_{\textbf{k}'}|\phi_{\textbf{k}'}|^2})$ and $\phi_\textbf{k} = \sin(k_xa)-2\sin(\frac{k_xa}{2})\cos(\frac{\sqrt{3}k_ya}{2})$ is the $f_x$-wave form factor. The corresponding $f$-wave VP order is given by  $\frac{1}{N_e}\sum_\textbf{k} \tilde{\zeta}_\textbf{k} \langle \psi^\dagger_\textbf{k} \tau_z \sigma_0 \psi_\textbf{k} \rangle$, where  $\tilde{\zeta}_\textbf{k}=\zeta_\textbf{k}/\sqrt{N_\textbf{k}^{-1}(\sum_{\textbf{k}'}|\zeta_{\textbf{k}'}|^2})$ and $\zeta_\textbf{k} = \sin(\sqrt{3}k_ya)-2\sin(\frac{\sqrt{3}k_ya}{2})\cos(\frac{3k_xa}{2})$ is the $f_y$-wave form factor. These order parameters are plotted in Figs.~\ref{Orderpara_IVC_phase}(a) and (b) as a function of the doping.

We thus see that the insulating KIVC state maintains its character on hole doping from $\nu=-2$. However, a $f$-wave TIVC order keeps increasing in the system with hole doping.

\subsection{Quantum phase transitions}

We now focus on the different symmetry-broken phases that appear within the HF theory as the carrier density is tuned from $\nu=-2$ to $\nu=-3$.  As we have mentioned before, the system remains in a predominantly KIVC state (with a subdominant $f$-wave TIVC order) at low doping near $\nu=-2$. The KIVC order parameter, plotted in Fig.~\ref{Phase_diagram_dominant_order}(a) as a function of carrier density, remains at a constant value of $\sim 0.95$ till $\nu= -2.48$, when it falls suddenly to zero. At the same filling fraction, the valley polarization (this is the uniform valley polarization with an $s$-wave form factor in contrast to the valley polarization with $f$-wave form factor discussed in the previous section) rises to $1$. This shows that the occupied HF bands now belong to a single valley instead of being a coherent superposition of states from the two valleys. The transition is first order and is similar to a spin-flop transition (albeit in valley space). Fig.~\ref{Phase_diagram_dominant_order}(a) shows the order parameters as a function of filling for a strongly interacting system with the gate distance  $d=100$ nm. The same order parameters are plotted for a system with $d=20$ nm in Fig.~\ref{Phase_diagram_dominant_order}(b). While decreasing the gate distance leads to larger screening from the gates and weaker Coulomb interactions, we find that the critical filling at which the transition happens does not vary with $d$. The phase diagram in the $\nu-d$ plane is shown in Fig.~\ref{Phase_diagram_dominant_order}(c), and we see that the transition line from KIVC to a VP state is almost independent of $d$.

The VP state obtained at the phase transition, is diagonal in the original non-interacting band basis; i.e. they are consist of an equal superposition of states lying on the $A$ and $B$ sublattices. As a result, the sublattice polarization is zero in this state. As the carrier density is further tuned, at $d=100$ nm, a sublattice polarization starts to build up beyond $\nu=-2.60$. This is the QAH state, which breaks inversion symmetry and leads to an anomalous Hall response due to the intrinsic Berry curvature of the bands. The sublattice polarization starts from zero, indicating a continuous phase transition in this case, as seen in Fig.~\ref{Phase_diagram_dominant_order}(a). The valley-polarized QAH state persists till $\nu=-3$, with the sublattice polarization increasing as the carrier density is tuned towards $\nu=-3$. 

The presence of the QAH phase transition, however is not a universal feature. In Fig.~\ref{Phase_diagram_dominant_order}(b),  we do not see the appearance of a QAH state at weak coupling ($d=20$ nm). Instead, beyond $\nu=-2.86$, the system seems to find a ground state which has all orders (KIVC, TIVC, VP, QAH). This mixed state is however, not robust, and details of this state depend on the subtraction schemes used in the calculation \cite{Supple}. This is not surprising since the state at $\nu=-3$ within the HF theory is known to depend on details of the regularization scheme \cite{Kwan_PRX_2021}. The phase diagram in the $\nu-d$ plane is shown in Fig.~\ref{Phase_diagram_dominant_order}(c). The $M$ or mixed phase near $\nu=-3$ is sensitive to details of subtraction scheme, while the rest of the phase diagram is robust with respect to these technical details.

In experiments \cite{Sherlin_Science_2020}, the QAH state is observed when the sample aligns with the hBN substrate, which induces different onsite potentials for $A$ and $B$ sublattices. This explicit symmetry breaking between the sublattices results in nonzero sublattice polarization. However, our results reveal that strong Coulomb interaction at large gate distances can spontaneously favour the QAH phase without explicit hBN alignment.

Another phase transition, which occurs in parallel, involves the spontaneous breaking of the $C_3$ symmetry of the lattice. This happens either in the VP phase or in the QAH phase depending on the value of $d$ ( see the $C_3$ invariant and broken phases in $\nu-d$ plane in Fig.~\ref{Phase_diagram_dominant_order}(d)). As the hole density increases, one of the hole pockets grows in size, and the $C_3$ symmetry breaking is correlated with this Fermi surface touching the edges of the mBZ.  To see this, we plot the Fermi surface contours in the two HF bands in the mBZ at $\nu=-2.72$ (Fig.~\ref{Phase_diagram_dominant_order}(e)), where the system is in the $C_3$ symmetric phase and $\nu=-2.74$ (Fig.~\ref{Phase_diagram_dominant_order}(f)), where the $C_3$ symmetry is broken. These plots are shown for $d=100$ nm, where the $C_3$ symmetry breaking occurs in the QAH phase. Note that the small Fermi surface in the other HF band maintains $C_3$ symmetry across the transition. Finally we would like to note that there are three degenerate solutions in the $C_3$ broken phase, with the Fermi contours rotated by $120^\circ$. We have explicitly checked this in our calculations. Depending on the details of the convergence process of the equations, one of these solutions is randomly chosen. 

We have thus provided a comprehensive phase diagram of the system within HF theory as a function of carrier density and interaction strength (parametrized by the gate distance). We find a robust KIVC to VP transition which is very weakly dependent on the interaction strength. We find a further transition into a QAH phase, but only at very strong interactions. The fate of the system near $\nu=-3$ at weak interactions is susceptible to details of numerics, while a robust QAH phase is seen at strong interactions.

\section{Electronic Structure in Different Quantum Phases}
The quantum phase transitions of the MATBG system as a function of the carrier density leave their signatures in the single-particle HF bands of the system. It is obvious that the bands in the different phases are built out of different combinations of spin, valley, and sublattice degrees of freedom, and we have discussed these details in the previous section. However, the quantum phases also leave their signatures in the band dispersions. There are two important features of band dispersions which are controlled by the quantum phases : (a) symmetries of the phase dictate certain degeneracies in the dispersion and (b) the transitions between phases lead to Lifshitz transitions in the resulting band dispersions. We will now discuss these issues for MATBG in the doping regime $-3 < \nu < -2$.

\subsection{Symmetries and degeneracies}
\begin{figure}
\vspace{2ex}
 \begin{center}
 \includegraphics[scale=0.45]{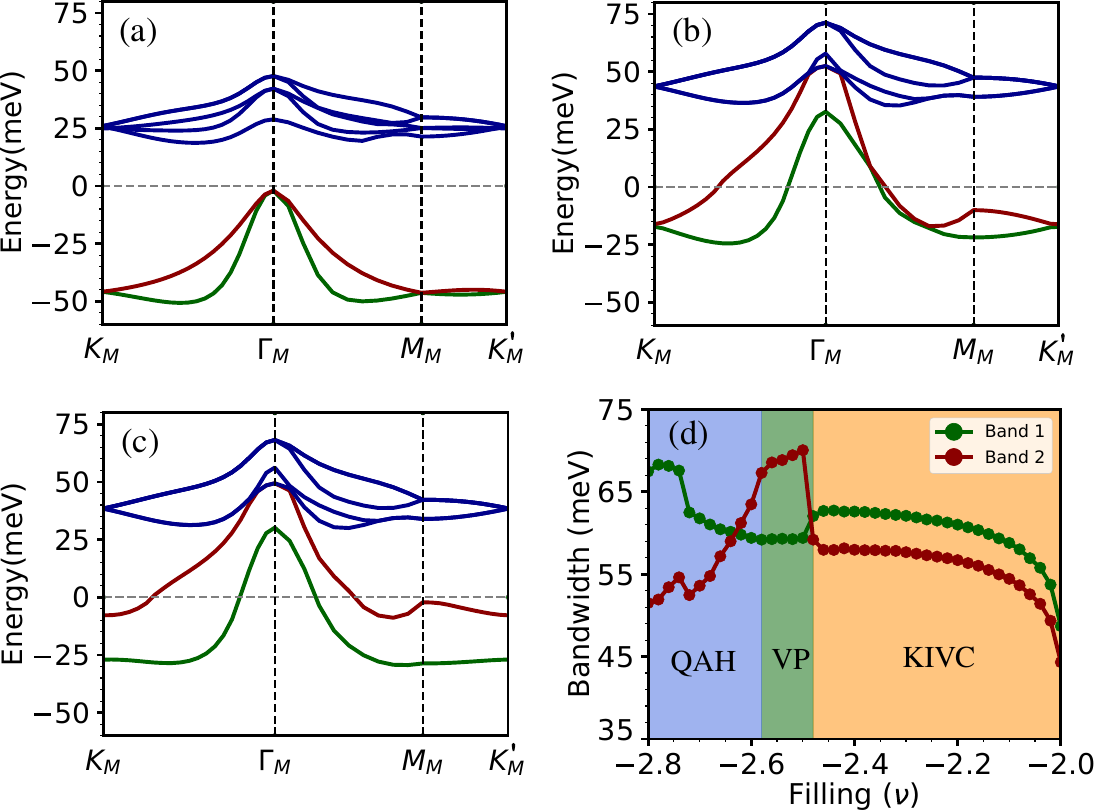}
 \end{center}
 \vspace{-3ex}
 \caption{The bandstructure and bandwidth at $d$ = 100 nm. (a)-(c) The bandstructure along the highsymmetry directions in the mBZ at $\nu$ = -2.00, -2.52, and -2.64, respectively. These fillings correspond to the IVC, VP, and QAH phase, respectively. The two filled bands are shown in green and red. (d) The bandwidths of the two filled bands in(a)-(c) as function of $\nu$.}
  \label{Bandstructure}
\end{figure}

We first consider the KIVC state at the insulating point $\nu=-2$. Out of the eight HF bands, six are unoccupied and are separated from the two occupied bands  by a band gap of $\sim 20$ meV. The band dispersions are shown in Fig.~\ref{Bandstructure}(a). Focusing on the two occupied bands, we find that the bands are degenerate at the $K_M$ and $K_M^{'}$ points, where there is a remnant of the Dirac cone inherited from the original graphene band structure. Additionally, the bands are degenerate at the $\Gamma_M$ and $M_M$ points. As a consequence, the two HF bands have almost identical bandwidths. The degeneracies at the $K_M$ and $K_M^{'}$ points are protected both by the $C_3$ symmetry as well as the $C_2T'$ symmetry which is present in this phase (Note that the obtained $s-$wave KIVC phase has order parameter $\tau_y\sigma_y$, which preserves the $C_2T'$ symmetry. In contrast, the KIVC phase with order parameter $\tau_x \sigma_y$ satisfies the $C_2T$ symmetry.). Additionally, the system also preserves the $T'$ symmetry, which leads to degeneracies at the $\Gamma_M$ and the $M_M$ points \cite{Kozii_PRB_2022}. 

As we dope the system, a hole pocket forms in these two bands around the $\Gamma_M$ point. The bandwidth of both the bands increases slightly with carrier doping, as seen in Fig.~\ref{Bandstructure}(d). However, the degeneracies, protected by the symmetries remain throughout this phase. It is interesting to note that as the carrier concentration is changed from $\nu=-2$ to $\nu=-3$, one expects one of the two bands to peel off and join with the six bands above. However, the symmetry protected band degeneracies will prevent that from happening. This is also an indicator that more symmetry breaking is likely to happen as the carrier densities are tuned in the system.

The system transitions to a valley polarized phase at $\nu$ = -2.48 through a first order phase transition. The band dispersion in this phase ( at $\nu=-2.52$ for $d=100$ nm) is shown in Fig.~\ref{Bandstructure}(b). The VP phase preserves the $C_2T'$ symmetry, while breaking the $T'$ symmetry. Additionally, around $d=100$ nm, the entire valley polarized phase preserves the $C_3$ symmetry, while $C_3$ is broken within the valley polarized phase at lower values of $d$. The presence of $C_2T'$ and $C_3$ symmetry protects the band degeneracies at $K_M$ and $K_M^{'}$ points. However, the degenracy at $\Gamma_M$ point is lifted due to breaking of the $T'$ symmetry. The top edge of one of the bands (around $\Gamma_M$ point) gets lifted up and gradually mixes with the other high energy bands, while that of the other one goes down due to mode coupling effects. This separation of the bands at $\Gamma_M$ point and the degeneracy at $K_M$ and $K_M^{'}$ points are reflected in the fact that bandwidth of one of the HF bands shows a jump to a higher value (and the other one jumps to a lower value) at the quantum phase transition.

At a higher value of hole doping around $\nu=-2.6$, the system develops a sublattice polarization and goes into a QAH state through a continuous phase transition. The QAH phase breaks $C_2T'$ symmetry and the degeneracy of the bands at $K_M$ and $K_M^{'}$ points are lifted. This can be seen in Fig.~\ref{Bandstructure}(c) where we plot the band dispersions of the system at $\nu=-2.64$ for $d=100$ nm. The bands then continue to peel off from each other as the carrier density is tuned further. We note that at low values $d<50$ nm, the $C_3$ symmetry is already broken in the VP phase and the degeneracy of the $K_M$ and $K_M^{'}$ are lifted, before the system develops the QAH order.

It is interesting to observe the evolution of the bandwidths through these transitions. As seen in Fig.~\ref{Bandstructure}(d), the band width of one of the bands starts to undergo a steady decline beyond the transition to the QAH phase, as the $K$ degeneracies are lifted and  the band gets pulled into the manifold of high energy bands. The additional jump around $\nu=-2.72$ is due to the reorganization resulting from $C_3$ symmetry breaking in the system.

\begin{figure*}[t]
\begin{center}
\includegraphics[scale=0.8]{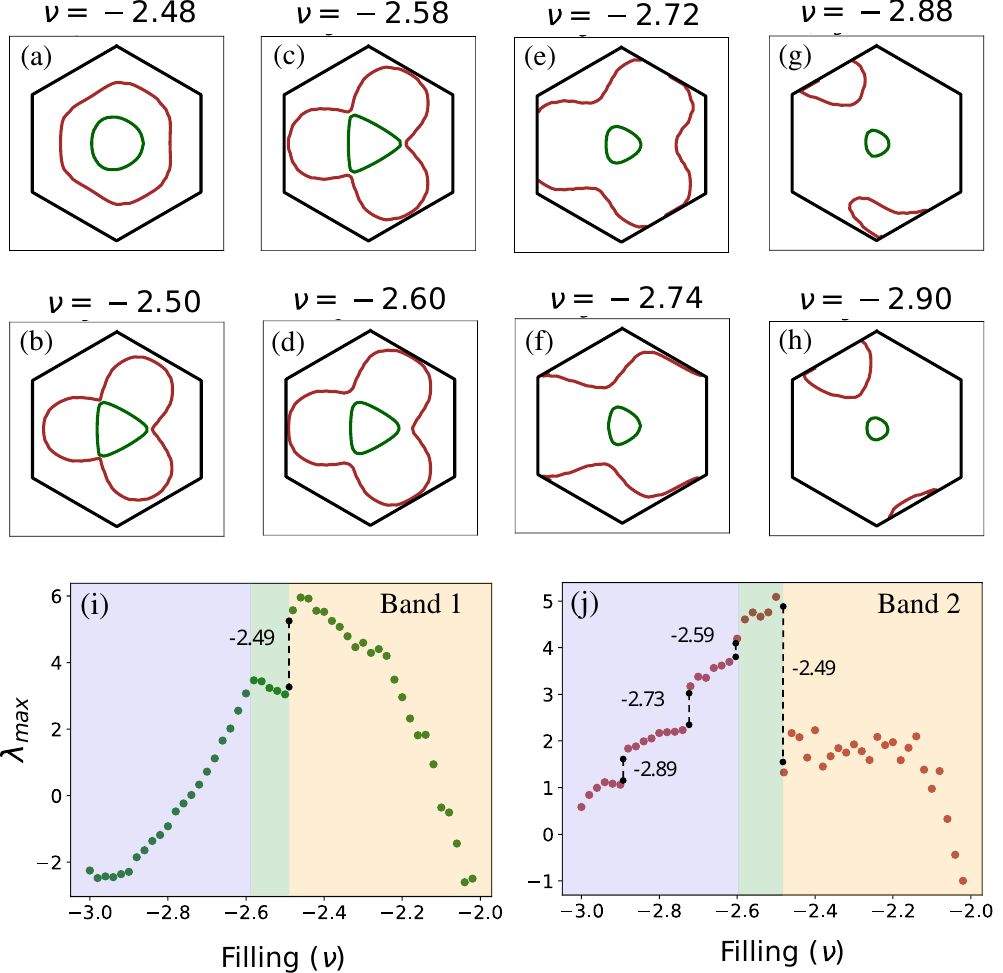}    
\end{center}
\caption{Lifshitz transition (LT): (a)-(h) Fermi surfaces of the lowest (green) and second-lowest (red) HF modified valence bands around the four prominent LT densities at $d$=100 nm. The sudden reconstruction of the HF Fermi surfaces can be seen across the LTs. The largest eigenvalue of the effective inverse mass tensor is line-integrated along the Fermi surface and has been plotted in (i) and (j) for the two bands labeled band 1 and band 2, respectively. In (i), the measure ($\lambda_{max}$) for the lowest band shows one clear discontinuity at $\nu=-2.49$ and one kink at $\nu=-2.59$. In (j), the measure for the second-lowest band shows four discontinuous jumps at the marked densities, tracking the LTs quantitatively. The backgrounds of (i) and (j) are shaded according to the corresponding symmetry-broken HF phases, highlighting that two of the four LTs coincide exactly with the phase-transition points.}
\label{Lifshitz_trans}
\end{figure*}

\subsection{Lifshitz transitions}

In the metallic density regime, transitions between distinct interaction-driven symmetry-broken states are accompanied by a reconstruction of the HF band structure. This reconstruction can change the topology of the Fermi surface, giving rise to Lifshitz transitions (LTs) at characteristic densities. We therefore track the LT in the HF bands as a complementary signature of the transitions between the symmetry-broken states. To see this, we plot the two Fermi surfaces in the two HF bands for different values of carrier densities in Figs.~\ref{Lifshitz_trans}(a)-(h). In the KIVC phase, we find concentric Fermi surfaces around the $\Gamma_M$ point, as seen in Fig.~\ref{Lifshitz_trans}(a). However, the nature of the Fermi surface changes when we move to the VP phase beyond $\nu=-2.48$. This is clearly seen in Fig.~\ref{Lifshitz_trans}(b) where the Fermi surfaces are plotted at $\nu=-2.50$, just in the VP phase. One of the Fermi surfaces show a large flower like topology, while the other forms a small triangle around the $\Gamma_M$ point.  These Fermi surfaces (which are in different bands) almost touch each other. We note that the dispersions of the bands also change across these transitions, as we discussed in the last section, but here we are focusing on the Fermi surface only. In Figs.~\ref{Lifshitz_trans}(c) and (d), we plot the Fermi surfaces just on either side of the transition to the QAH phase. We see that there is another Lifshitz transition where the two Fermi surfaces now separate out. As we increase hole doping, we see another transition around $\nu=-2.72$. This is the $C_3$ symmetry breaking transition, and the corresponding transition in the Fermi surface topology is seen in Figs.~\ref{Lifshitz_trans}(e) and (f). The $C_3$ symmetric closed Fermi surface changes to an open Fermi surface with only $C_2$ symmetry. Note that the changes are happening in one of the bands, while the other Fermi surface is evolving smoothly. Figs.~\ref{Lifshitz_trans}(g) and (h) show how this open Fermi surface evolves on further doping the system.

Beyond the visual representation, we can quantify these transitions by considering the curvature of the HF bands. For each band, we first construct the inverse mass matrices at each momentum point ($k_x$,$k_y$) 

\begin{equation}
    \mathcal{M}_i^{-1}(k_x,k_y)=\begin{bmatrix}
        \frac{d^2\varepsilon_i}{dk_x^2}&\frac{d^2\varepsilon_i}{dk_xdk_y}\\
        \frac{d^2\varepsilon_i}{dk_ydk_x}&\frac{d^2\varepsilon_i}{dk_y^2}\\
    \end{bmatrix}.
\end{equation}

where $\varepsilon_i(\textbf{k})$ is the dispersion of the HF band. We diagonalize this matrix and consider the largest eigenvalue of this matrix $\lambda_{max}(k_x,k_y)$. This quantity is then integrated along all points on the Fermi surface. In Figs.~\ref{Lifshitz_trans}(i) and (j), we plot this Fermi surface integrated largest eigenvalue for the two bands as a function of the carrier density. We find that both bands show a discontinuity in this quantity at the KIVC-VP phase transition. At the transition from the VP to the QAH phase, one of the bands show a jump discontinuity, while the other band shows a weaker derivative discontinuity. The other jumps related to the $C_3$ symmetry breaking only show up in one of the bands which undergo the closed to open Lifshitz transition. The inverse mass tensor and its eigenvalues play an important role in thermal transport in metallic systems, and one would expect low temperature heat transport to pick up signatures of these phase transitions.

\section{Conclusion}

In conclusion, we studied the ground state phase of the normal metallic state of the MATBG system in the filling range of $\nu=-2$ to $\nu=-3$. We found that the IVC phase near $\nu=-2$, possesses the $s-$wave KIVC and $f-$wave TIVC orders. Although the KIVC order dominates the IVC phase, the strength of the TIVC order reaches a significant fraction of the strength of the KIVC order. This finding establishes that the TIVC order can be stabilized in the unstrained samples without the electron-phonon coupling, as had been proposed in previous theoretical studies.

We further found that for fillings near $\nu=-2$ the phases show a weak dependence on the gate distance, while close to $\nu=-3$, we found a stronger dependence on the gate distance. The IVC phase is found to be the ground state close to $\nu=-2$ at all gate distances. The system shows two phase transitions at large gate distances. The first transition from KIVC to VP is a first order like phase transition while the second transition from VP to QAH is a second order like phase transition. The study also reveals that at $\nu=-3$, the system prefers to be in the QAH phase at large gate distances when the Coulomb interaction is strong without the need of alignment with the hBN substrate. At low gate distances, near $\nu=-3$, we obtain a mixed phase state.

We believe that our study of the normal state phase diagram will be further useful in the investigation of superconductivity, as recent experiments strongly suggest that the superconductivity in this system could arise as an instability of the underlying metallic ground state.

\section{Acknowledgment}
This work is supported by the Department of Atomic Energy, Government of India, under Project Identification Number RTI-4012 and RTI-4013. Computations were carried out on the computing clusters at the Department of Theoretical Physics, TIFR, Mumbai, and at the Academic Center for Materials and Nanotechnology, AGH University of Krakow, Poland.  We would also like to thank Ajay Salve and Kapil Ghadiali for computational support. UG would like to thank Dr. Debasmita Giri for useful discussions related to Hartree-Fock calculations in \moire materials. 

\end{document}